\documentclass[onecolumn,preprint,preprintnumbers,amsmath,amssymb]{revtex4}

\usepackage{graphicx}
\usepackage{dcolumn}
\usepackage{bm}

\begin{document}

\title{Under which conditions is quantum brownian motion observable in a microscope?}

\author{L.E. Helseth}

\affiliation{Department of Physics and Technology, University of Bergen, N-5007 Bergen, Norway}%

\begin{abstract}
We investigate under which conditions we can expect to observe quantum brownian motion in a microscope. Using the fluctuation-dissipation theorem, we investigate quantum brownian motion in an ohmic bath, and estimate temporal and spatial accuracy required to observe a crossover from classical to quantum behavior.
\end{abstract}

\pacs{Valid PACS appear here}
\maketitle

\section{Introduction}
Classical fluctuations play important roles in both understanding everyday phenomena as well as designing new sensitive detections systems, with the necessity of a deeper understanding of detection limits being a driving factor for research on nanosystems over the last decades\cite{Russel,Ekincia}. As one develops more sensitive systems allowing detection of smaller and smaller fluctuations, the quantum regime is ultimately approached. Quantum Brownian motion has been explored theoretically for decades\cite{Agarwal,Ford,Ford1,Taylor}, but direct observations have so far remained elusive. Indirect detection of quantum fluctuations relies on interpretation of spectral noise characteristics, where a crossover to quantum behavior is seen at low temperatures\cite{Koch}. Recent research on nanomechanical resonators at millikelvin temperatures has demonstrated position detection roughly an order of magnitude larger than the standard quantum limit $\Delta x _{SQL} =\sqrt {\hbar/2m\omega _0 }$\cite{LaHaye}. Here $h=2\pi \hbar$ is Planck's constant, $m$ is the mass and $\omega _0$ is the resonance frequency of the resonator, such that a system of mass $10^{-15}$ $kg$ oscillating at a frequency in the 0.1 GHz range has $\Delta x _{SQL} \sim 10^{-14}$ m. Due to back-action of the measurement system the smallest achievable detection limit is slightly larger\cite{Caves}, but of the same order of magnitude as the estimate above.

Designing experiments which allow direct imaging of quantum fluctuations is by no means an easy feat. Conventional microscopes optical light microscopes can typically only resolve objects larger or comparable to the wavelength of radiation. However, small displacements can be measured with much higher accuracy due to the rapid progress in modern optical tracking techniques. In fact, tracking of single molecules have revealed a whole new world of non-equilibrium mechanics with nanometer accuracy\cite{Crocker,Mason,Yildiz}. However, so far particle tracking techniques have been mainly limited to the classical physics domain, despite the fact that the displacement resolution is rapidly marching towards the $10^{-14}$ m required to approach the standard quantum limit above. This naturally leads one to ask how far such techniques can go, and how small fluctuations we would be able to observe. The current letter is an attempt to predict under which conditions we will be able to see quantum brownian motion of particles in a conventional microscope.

\section{Displacement resolution of the imaging system}
Consider a particle located at the focal plane of a microscope's objective. The scattered wave from the particle is collected by an objective lens of numerical aperture NA, and then imaged through the microscope onto an imaging device. The wavelength (in vacuum) is $\lambda _0$ and the power incident on the detector is $P_0$. The microscope may use microwave, infrared, visible or X-ray radiation, thus posing no restrictions on the wavelength at this point. A particle tracking algorithm relies on correctly identifying the position of the particle by using the relationship between the detected signal current and position $x$.
As a result of inherent white noise in the detection system, the signal power is distributed around the mean value $P_0$ even in absence of the scatterer such that the standard deviation $P_n$ represents the noise power\cite{Helseth}. The electronic noise and the actual position fluctuations are uncorrelated. Thus, if the signal noise is sufficiently large, it may under certain circumstances appear as a particle is performing Brownian motion even in cases where it is completely fixed\cite{Helseth,Helseth2}. Thus, when studying position fluctuations one must always account for the apparent diffusion coefficient due to electronic noise. It was found in Ref. \cite{Helseth} that the apparent fluctuations about $x=0$ can be given as
\begin{equation}
\Delta x _a  \approx \frac{\lambda_0}{NA} \frac{0.2}{\sqrt{SNR}} \,\,\, ,
\label{squ}
\end{equation}
where the signal to noise ratio is given by $SNR=P_0 /P_n$. An image system monitoring particle displacements cannot provide a displacement resolution significantly better than the noise-induced apparent fluctuations. That is, it is reasonable to take $\Delta x_a$ as a measure of the smallest achievable displacement resolution. In a good microscope we may have $\lambda_0 =0.55$ $\mu m$, $NA\approx 1$ and $SNR \approx 10000$, which gives $\Delta x_a \approx 1$ $nm$. The nanometer mark is routinely achieved in current commercial and home-built microscopes\cite{Crocker,Mason,Yildiz}, and it is expected that future systems will be able to resolve displacements that are orders of magnitude smaller\cite{Helseth}. It should be pointed out that Eq. \ref{squ} represents the apparent displacement due to noise, and thus we must require that the actual displacement due to diffusion is larger than $\Delta x_a$. Digitizing the images introduces additional sources of error, but this is will not be considered here, since such errors may in fact be reduced by selecting a very high resolution imaging system.

\section{Quantum brownian motion}
Utilizing the good signal to noise ratio of modern microscopes, scientists have been able to track walking motor proteins with nanometer resolution\cite{Yildiz}. Such molecules are nonetheless operating in the classical regime. As the displacement resolution is steadily increasing, the ultimate goal would be to track a particle or molecule in the quantum regime. However, first one needs to understand how the systems behaves in this regime. In order to model quantum fluctuations, a proper analysis would require an approach where the particle is connected to a bath. Since quantum dissipative processes typically are non-Markovian, the resulting colored noise spectrum results in anomalous diffusion. Several different approaches to solve the  problem exists, based on e.g. the master equation or the quantum Langevin equation\cite{Ford,Ford1,Taylor}. We will here use the approach of Ford et al. \cite{Ford,Ford1} since this allows us to estimate the diffusion coefficient of linear systems rather straightforwardly in a similar manner as is done for classical fluctuations.

The idea is to consider a particle that is exposed to a time-varying force $F(t)$ such that the position operator $x(t)$ can be described by the susceptibility $\alpha$ through the following equation
\begin{equation}
x(t) =\int_{0}^{\infty} \alpha (\tau) F(t-\tau) d\tau \,\,\, .
\label{AA}
\end{equation}
According to the fluctuation-dissipation theorem, first formulated by Callen and Welton\cite{Callen}, the symmetrized correlation is given by
\begin{equation}
\frac{1}{2}\left< x(t)x(t+\tau) + x(t+\tau)x(t) \right> =\frac{\hbar}{\pi} \int_{0}^{\infty} coth\left ( \frac{\hbar \omega}{2k_B T}\right) Im\left[ \alpha(\omega)\right] cos(\omega \tau) d\omega \,\,\, .
\label{AB}
\end{equation}
where $Im[\alpha]$ is the imaginary part of the susceptibility, $k_B$ is Boltzmanns constant and $T$ the temperature of the bath. If we are not interested in the time-development, we may set $\tau =0$ such that
\begin{equation}
\left< x^{2} \right> =\frac{\hbar}{\pi} \int_{0}^{\infty} coth\left ( \frac{\hbar \omega}{2k_B T}\right) Im\left[ \alpha(\omega)\right] d\omega \,\,\, .
\label{ABE}
\end{equation}
The susceptibility for a general non-Markovian system exhibiting memory effects can be found using the quantum Langevin equation on the form\cite{Ford,Ford1}
\begin{equation}
m\ddot{x}(t) + \int_{-\infty}^{t} \mu (t -t')\dot{x}(t')dt' + \frac{dV(x)}{dx} =F(t) \,\,\, ,
\label{Langevin}
\end{equation}
where $\mu (t)$ is the memory function and $-dV(x)/dx$ is the external force.
In the current study we will assume that the particle is immersed in an ohmic heat bath experiencing a friction coefficient $\gamma$. We then have $\mu (t) =\gamma \delta (t)$, which is a model that can be used to explain the experimental results of a classical fluctuations at relatively long time scales. Under such conditions, the particle state changes so slowly that the surrounding bath has time to re-equilibrate. However, for small time-intervals this may not be the case when the particles experience non-uniform external forces\cite{Zhu}. Of perhaps even greater interest is the time-regime where we can expect crossover from classical to quantum brownian motion. Although it is not yet clear whether an ohmic bath can be realized experimentally in the quantum regime, we believe the results coming out of such an analysis will give reasonable estimates of the time and spatial scales involved. Moreover, it is known that baths containing, e.g. trapped ions, can be engineered\cite{Myatt}, thus bringing hope that an ohmic bath is a realistic scenario. It should also be pointed out that we neglect the contribution of the thermal radiation to the bath, and therefore assume that the light scattered off the particle does not change the characteristics of the bath given below (for a detailed studied of particles immersed in a strongly coupled blackbody radiation bath, see Ref. \cite{Ford1}). Thus, the frequency of the light observed in the optical microscope is given by $\omega _{l} =2\pi c/\lambda _0$, and is not the same as the frequencies of radiation modes associated with the thermal bath.

In addition to the friction force, let us now assume that the particle experiences a harmonic external potential, $V(x)=1/2m\omega _0 ^{2} x^{2}$, where $\omega _0$ is the resonance frequency. Such an external force may come about if the particle is trapped by, e.g., an electromagnetic field. In an ohmic bath, the susceptibility can be found from Eq. \ref{Langevin} to be
\begin{equation}
\alpha (\omega) =\frac{1}{m}\frac{1}{\omega _0 ^{2} -\omega ^{2} -i\frac{\gamma}{m}\omega} \,\,\, .
\end{equation}
For ease of derivation of the following results it is useful to write the susceptibility on the form
\begin{equation}
\alpha (\omega) =  \frac{1}{2m\sqrt{\omega _0 ^{2} -\left( \frac{\gamma}{2m} \right) ^{2} }} \left[ \frac{1}{\sqrt{\omega _0 ^{2} -\left( \frac{\gamma}{2m} \right) ^{2} } -\omega -i\frac{\gamma}{2m}} + \frac{1}{\sqrt{\omega _0 ^{2} -\left( \frac{\gamma}{2m} \right) ^{2} } +\omega + i\frac{\gamma}{2m}} \right]  \,\,\, .
\label{alpha1}
\end{equation}

In the limit $\gamma /2m \rightarrow 0$, it is straightforward to see from Eqs. \ref{ABE} and \ref{alpha1} that\cite{Li}
\begin{equation}
\left< x^{2} \right> =\frac{\hbar}{2m\omega _0} coth\left ( \frac{\hbar \omega _0}{2k_B T}\right)  \,\,\, .
\label{ABEE}
\end{equation}
Equation \ref{ABEE} is in the high-temperature limit given by
\begin{equation}
\left< x^{2} \right> \approx \frac{k_B T}{m\omega _0 ^{2}} \,\,\, , \,\,\,  \frac{\hbar \omega }{k_B T} \ll 1 \,\,\, .
\label{ABEE1}
\end{equation}
In the low-temperature regime we have the standard quantum limit
\begin{equation}
\left< x^{2} \right> = \left( \Delta x_{SQL} \right) ^{2} \approx \frac{\hbar}{2m\omega _0} \,\,\, , \,\,\,  \frac{\hbar \omega }{k_B T} \gg 1 \,\,\, .
\label{ABEE2}
\end{equation}
It should be pointed out that in order to reach the standard quantum limit, a position accuracy of $<10^{-14}$ $m$ is required if one aims at tracking a particle of mass $10^{-15}$ $kg$ at a resonance frequency of 0.1 GHz. According to Eq. \ref{squ} this would require $SNR \sim 10^{14}$, which is orders of magnitude from that achieved with current microscopy techniques. The required force constant in this case is rather large ($k\approx 400$ $N/m$), although it could in an electromagnetic trap be as low as $k\sim 10^{-9}$ $N/m$\cite{Helseth3}. For such weak traps one thus finds $\omega _0 \sim 1000$ $s^{-1}$ and $\Delta x _{SQL} \sim 10^{-11}$ m. Smaller particles and single molecules are expected to have a mass $m \leq 10^{-20}$ kg. This reduces the standard quantum limit to $\Delta x _{SQL} \sim 10^{-9}$ m, and the required signal to noise ratio is now $\sim 10^{4}$, which is within the range of modern microscopy techniques. At this point it should be emphasized that position sensitivity better than $10^{-15}$ m has been achieved using optical cavities\cite{Numata}, but such setups have up to now not allowed one to track particles and therefore cannot be applied in the current context.

\begin{figure}
\includegraphics[width=12cm]{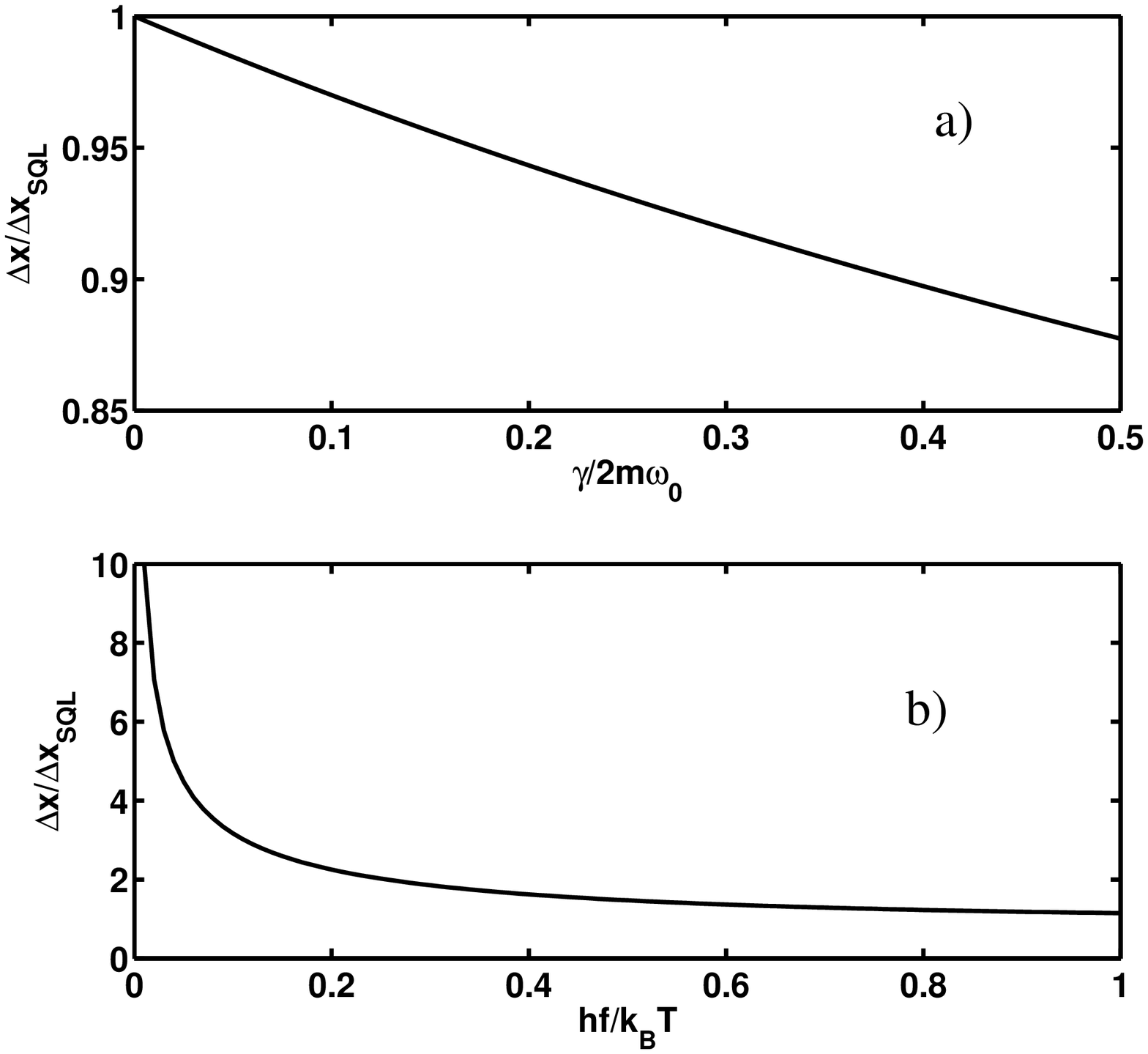}
\caption{\label{f1} The relative position accuracy $\Delta x /\Delta x_{SQL}$ displayed as a function
of $\gamma /2m$ for $hf/k_BT =0$ (a), and $hf/k_BT$ for $\gamma /2m =0$ (b).}
\vspace{2cm}
\end{figure}

The standard quantum limit should not be applied carelessly when studying particles performing quantum brownian motion, since it neglects dissipation. The dissipation can be accounted for in the low-temperature regime by using Eqs. \ref{ABE} and \ref{alpha1} (see also Refs. \cite{Caldeira,Li})
\begin{equation}
\left< x^{2} \right> =\frac{\hbar}{\pi m \sqrt{\omega _0 ^{2} -\left( \frac{\gamma}{2m} \right) ^{2} } } tan^{-1} \left[ \frac{\sqrt{\omega _0 ^{2} -\left( \frac{\gamma}{2m} \right) ^{2} }}{\frac{\gamma}{2m}} \right] \,\,\, , \,\,\,  \frac{\hbar \omega }{k_B T} \gg 1 \,\,\, .
\label{ABElowdiss}
\end{equation}
We have here assumed that $\omega _0 \geq \gamma/2m$. Figure \ref{f1} displays the relative position accuracy as a function of $\gamma /2m\omega _0$ for $\hbar \omega /k_BT \rightarrow \infty$ (a), and $\hbar \omega /k_BT$ for $\gamma /2m =0$ (b). We note that the fluctuations increase with temperature and decrease with the friction coefficient. This latter result suggests that a system with friction will experience smaller fluctuations than a non-dissipative quantum harmonic oscillator in its ground state. However, we found that that as long as $\hbar \omega /k_BT > 2$ and $\gamma /2m\omega _0 < 0.5$, the deviations from the standard quantum limit are less than 20$\%$. Thus, as long as we stay within these limits the conclusions for an ideal oscillator may directly be adopted to a dissipative oscillator as well. Combining eqs. \ref{squ} and \ref{ABEE2} we find that in order to be able to resolve quantum fluctuations the signal-to-noise ratio must be
\begin{equation}
SNR \geq \frac{0.1m\omega _0 \lambda _0 ^{2}}{\hbar NA^{2}} \,\,\, .
\end{equation}

Above we have considered a bound particle in an harmonic potential. In such cases it makes good sense to evaluate $\left< x^{2} \right>$. However, if the particle is not bound by an harmonic potential, but is instead free to move within the ohmic bath, one is mostly interested in the time-evolution. In order to study the time-evolution of the fluctuations, one must find the mean square deviation (MSD), which was written by Ford and O'Connell as\cite{Ford1}
\begin{equation}
M(\tau)=\left< \left[x(t) - x(0) \right] ^{2} \right> = \frac{2\hbar}{\pi} \int_{0}^{\infty} coth\left( \frac{\hbar \omega}{2k_B T}\right) Im\left[ \alpha(\omega)\right] \left[ 1-cos(\omega \tau) \right] d\omega \,\,\, .
\label{AC}
\end{equation}
The time-derivative of the MSD, $S(\tau)=dM(\tau)/d\tau$ is particularly useful, since for normal classical diffusion we have $M(t)=2Dt$ such that $S(\tau )=2D$, i.e. given directly by the diffusion coefficient $D$. If $dM/dt$ is time-dependent, the system is said to exhibit anomalous diffusion.
In the case of a free particle ($\omega _0 =0$) experiencing a friction coefficient $\gamma$, the diffusion coefficient is given by\cite{Ford1}
\begin{equation}
D(\tau)= \frac{2\hbar \gamma}{\pi m^{2}} \int_{0}^{\infty} \frac{1}{\omega ^{2} +\left( \frac{\gamma}{m} \right)^{2} } coth\left( \frac{\hbar \omega}{2k_B T}\right) sin(\omega \tau) d\omega \,\,\, .
\label{Ad}
\end{equation}
Let us now assume that $\omega \ll \gamma/m$, such that the typical time-interval is $\gg m/\gamma$. With this we assume that the ohmic bath is engineered to have an upper cut-off frequency comparable to the natural decay rate of the system. In the high-temperature regime we then have
\begin{equation}
D(\tau) \approx  \frac{2k_B T}{\gamma}\,\,\, , \,\,\, \frac{\hbar \omega }{k_B T} \ll 1 \,\,\, ,
\label{Att}
\end{equation}
which has been observed experimentally in numerous classical microscopic systems (see e.g. Ref. \cite{Helseth2} and references therein). In the low-temperature regime we have\cite{Ford1}
\begin{equation}
D(\tau) \approx  \frac{2\hbar}{\pi \gamma \tau }\,\,\, , \,\,\, \frac{\hbar \omega }{k_B T} \gg 1 \,\,\, .
\label{Ad}
\end{equation}
The diffusion coefficients for the high and low-temperature regimes have the same order of magnitude when $\tau \sim \hbar /\pi k_B T$. Assuming that the thermal bath holds a temperature $T\sim 10$ K we find that $\tau \sim 10^{-13}s$, which means that we must measure the diffusion coefficient at very small time-intervals in order to see the crossover from quantum to classical fluctuations. That quantum effects take place at short time scales is not surprising, and could have been anticipated directly from Heisenbergs uncertainty relationship. That is, since uncertainty in energy and time are related according to $\Delta E \Delta t \sim \hbar$, we see that energy uncertainties larger than $k_B T$ gives rise to $\Delta t < \hbar /k_B T$. In order to be able to observe a crossover from classical to quantum brownian motion, we would like to be able to observe the diffusion at time scales given by $m/\gamma \ll \tau < \hbar /\pi k_B T$. Modern lasers are able to provide pulses of temporal duration of about one femtosecond, which could be used as the lower time scale of a typical experiment. Since we require that $\tau \gg m/\gamma $, it is seen that a molecule of mass $m\sim 10^{-24}$ $kg$ must experience a friction coefficient $\gamma \gg 10^{-9}$ $Ns/m$ within our approximation. Under these conditions one must be able to measure diffusion coefficients considerably smaller than $10^{-12}$ $m^{2}/s$ over a time interval from $10^{-13} - 10^{-15}$ $s$. Since the diffusion coefficient can be estimated from experimental data using $\Delta x ^{2}/2\Delta t$, we find that the smallest displacement we must be able to track is $\Delta x \sim \sqrt{2D\Delta t} \approx 4 \times 10^{-14}$ $m$. As we found above, this would require a very high signal to noise-ratio ($SNR \sim 10^{14}$). It is clear that further advances in using, e.g., laser-based scattering and particle tracking\cite{Mason,Helseth} is needed in order to bring the signal to noise ratio up in a conventional microscope if our aim is to observe a crossover from classical to quantum fluctuations.

\section{Conclusion}
We have investigated the conditions required for quantum fluctuations to be observable in a conventional microscope. The position of a molecule or particle can be observed in a microscope, thus making this quantity particularly interesting for studying the crossover from classical to quantum fluctuations. Quantities which are only indirectly observed through other properties, such as e.g. temperature fluctuations in nanosystems, would also be of interest. To this end, it has been suggested by Balatsky and Zhu that a quantum temperature fluctuations should be observable below the temperature $\hbar /k_B \tau _r$, where $\tau _r$ is the thermal relaxation time of the system\cite{Balatsky}. For small relaxation times, such as those found in nanosystems, one may expect the crossover temperature to be sufficiently large thus allowing direct observation of the temperature dependence of noise.

It should also be pointed out that in the current study only second moments have been considered, mainly due to the fact that these are most often reported experimentally. However, higher moments may be more sensitive to the crossover from classical to quantum fluctuations, but such a study is outside the scope of the current work.

\newpage

\newpage

\end{document}